\documentclass{article}
\usepackage[table]{xcolor}
\usepackage{spconf,amsmath,amssymb,graphicx}
\usepackage{booktabs,array,tabularx,capt-of,balance}
\usepackage{tikz,pgfplots}
\usetikzlibrary{positioning,fit}
\usepackage{microtype}
\usepackage[hidelinks]{hyperref}
\pgfplotsset{compat=1.18}

\hypersetup{pdftitle={AdaLoop: Adaptive-Depth Latent Reasoning for Audio Language Models}}
\newcommand{\best}[1]{\textbf{#1}}
\newcommand{\second}[1]{\underline{#1}}
\newcolumntype{N}{>{\centering\arraybackslash}p{3.3em}}
\newcolumntype{A}{>{\centering\arraybackslash}p{4.2em}}
\newcommand{\paperfont}{\fontsize{9}{10.8}\selectfont}
\definecolor{rowgray}{gray}{0.92}
\definecolor{headgray}{gray}{0.85}
\definecolor{tblNavy}{HTML}{243B53}
\definecolor{tblBand}{HTML}{EEF2F5}
\definecolor{tblStripe}{HTML}{F6F8FA}
\definecolor{tblOurA}{HTML}{E9F4F1}
\definecolor{tblOurB}{HTML}{EBF6F2}
\definecolor{tblOurInk}{HTML}{174E45}
\definecolor{tblRule}{HTML}{52687A}
\definecolor{tblGain}{HTML}{157951}
\newcommand{\hd}[1]{\textcolor{tblNavy}{\textbf{#1}}}
\newcommand{\grp}[1]{\textcolor{tblNavy}{\textit{#1}}}
\newcommand{\ours}[1]{\textcolor{tblOurInk}{\textbf{#1}}}
\newcommand{\gain}[1]{{\fontsize{7.2}{7.2}\selectfont\color{tblGain}$+$#1}}
\pgfplotsset{paperaxis/.style={
  width=0.475\textwidth,height=3.3cm,
  label style={font=\paperfont},tick label style={font=\paperfont},
  legend style={font=\paperfont,draw=none,fill=white,cells={anchor=west}},
  axis line style={black!65},tick style={black!65},
  grid=major,grid style={black!10},
  every axis plot/.append style={line width=1pt,mark size=1.8pt},
  scaled ticks=false}}

\title{AdaLoop: Adaptive-Depth Latent Reasoning\\for Audio Language Models}
\name{\begin{tabular}[t]{@{}c@{}}
Lee Seung-woo\textsuperscript{1}, Bowen Qi\textsuperscript{2}
\end{tabular}}
\address{\textsuperscript{1}Seoul National University\quad%
\textsuperscript{2}Shanghai Jiao Tong University}

\begin{document}
\ninept
\maketitle

\begin{abstract}
Large audio language models answer questions about speech, sound, and music,
yet their accuracy drops sharply on tasks that need fine-grained acoustic
analysis. Judging which of two speakers has the higher pitch demands iterative
signal-level reasoning that a content question does not. Current models spend
the same computational depth on both. We introduce \textbf{AdaLoop}, a
lightweight recurrent module that learns how many latent refinement steps a
given audio--question pair requires. A shared transformer block iterates over
the audio representation, guided by the question, while a learned halting
mechanism exits the loop once the representation is ready. AdaLoop adds fewer
than 3\% of the base model's parameters and plugs into any audio encoder--language
model pair without modifying either component. Evaluated on three architecturally
distinct models across MMSU, MMAU-Pro, and MMAR, AdaLoop raises the average
accuracy by 2.9 to 3.8 points, with the largest gains on perception-heavy
subtasks where the model learns to apply deeper reasoning.
\end{abstract}


\section{Introduction}
\label{sec:intro}
\vspace{-5pt}

Large audio language models (LALMs) unify speech, sound, and music
understanding through a single text
interface~\cite{qwen25,mimo,kimi,salmonn,ltuas,qwenaudio,speechgpt}, yet
their abilities are sharply uneven. Recent benchmarks that separate perception
from reasoning show that models understand meaning far better than they
perceive acoustic details~\cite{mmsu,mmaupro,mmar}; complementary evaluations
confirm this pattern across open-ended, spatial, and compositional
queries~\cite{audiobench,airbench,dynamicsuperb,mmau}. A model that transcribes a
sentence flawlessly may still fail to say which of two speakers sounds higher,
how many distinct sounds overlap in a scene, or whether a melody slowed down.
This perception gap limits applications that depend on fine-grained acoustic
judgment---paralinguistic analysis~\cite{wang2026parabridge}, nonverbal
vocalization understanding~\cite{ni2026nv}, and safety and privacy evaluation
of spoken interactions~\cite{wang2026voxsafebench,wang2026voxprivacy} all
require the model to reason about \emph{how} something sounds, not only
\emph{what} is said.

Several directions address this gap by enriching the training signal.
EvoAudio~\cite{wang2026evoaudio} builds a recursive self-improvement loop that
generates progressively harder acoustic training questions. Curriculum- and
reward-based approaches~\cite{curriculumlearning,selfpaced,tscl}
order synthetic questions from easy to hard or adapt
rewards as model responses evolve~\cite{af2,sari,audiozero,audiorubrics},
while self-training~\cite{star,selfinstruct}, test-time
reinforcement~\cite{aqattrl}, and pseudo-label scoring~\cite{sisda} offer
complementary paths. On the modelling
side, MimicLM~\cite{feng2026mimiclm} shows that fine-grained voice
characteristics can be captured autoregressively~\cite{valle,soundstream},
and Spatial-Omni~\cite{zhu2026spatial} extends perception to spatial audio
cues. Audio encoders---self-supervised~\cite{hubert,wavlm,wav2vec2},
supervised~\cite{whisper,ast,panns,conformer}, and
contrastive~\cite{clap,beats}---have also grown stronger, yet the
perception--reasoning imbalance persists.
These methods improve \emph{what the model can learn from}, but they do not
change how the model reasons at inference time: regardless of whether a
question is simple or difficult, the forward pass runs the same number of
layers with the same representation.

A separate line of work tackles inference-time reasoning depth.
RecurTrace~\cite{wang2026recurtrace} introduces loop-time memory that lets a
language model revisit and refine its latent representations over multiple
passes. AURAL~\cite{wang2026aural} applies adaptive latent reasoning with
chunked attention to speech, and Thinking-in-Depth~\cite{cai2026thinking}
extends recurrent latent reasoning to paralinguistically grounded spoken
dialogue. CoRELoop~\cite{feng2026coreloop} demonstrates that parameter-efficient
controlled recurrent refinement yields strong gains for audio deepfake detection,
a task where subtle spectral artifacts require iterative scrutiny. Each of these
methods shows that iterative refinement suits audio, but each targets a single
task and applies a fixed number of iterations. Generalist audio language
models~\cite{audioflamingo,pengi,gama,wavllm,spiritlm,miniomni,nexttgpt,macawllm}
still process every query at a single, fixed depth.

We propose \textbf{AdaLoop}, a module that combines recurrent latent refinement
with adaptive depth selection. Unlike fixed-loop approaches, AdaLoop predicts
how many refinement iterations a given audio--question pair requires and halts
early when shallow processing suffices. The depth is conditioned on both the
question and the audio, so the same model uses one or two passes for language
identification but six or more for a fine pitch comparison. The idea that
different inputs warrant different computational effort dates back to adaptive
computation time~\cite{graves2016adaptive,dehghani2019universal,pondernet,calm,mod,earlyexit,skipnet} and has proven
useful across sequential prediction tasks with multi-scale temporal
structure~\cite{yong2025predicting}. The transferability of structured
reasoning across modalities~\cite{wang2026exploring,wang2025towards} further
motivates our design: a single recurrent block that adapts its depth can span
the range from quick semantic checks to careful signal-level analysis without
task-specific engineering.

AdaLoop is lightweight (fewer than 3\% of parameters) and
architecture-agnostic: it plugs into the connector between any audio encoder
and language backbone without modifying either. We evaluate on
Qwen2.5-Omni-7B, Kimi-Audio-7B-Instruct, and MiMo-Audio-7B-Instruct across
MMSU, MMAU-Pro, and MMAR. AdaLoop raises the average accuracy by 2.9 to 3.8
points on every backbone, and the learned depth distributions confirm that the
model allocates deeper reasoning to perception-heavy tasks.

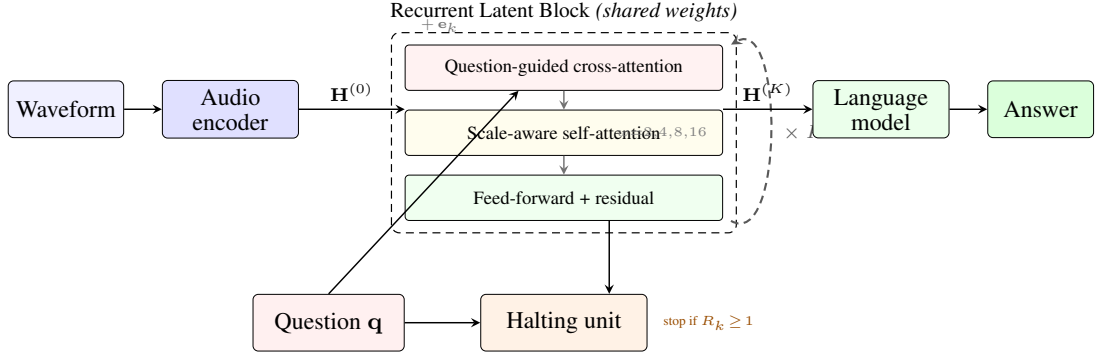
\begin{figure*}[t]
\centering
\begin{tikzpicture}[
  node distance=4mm and 5mm, >=stealth,
  box/.style={draw,rounded corners=2pt,minimum height=7.5mm,
    align=center,font=\paperfont\small,line width=0.35pt},
  inner/.style={draw,rounded corners=1.5pt,minimum height=6mm,
    align=center,font=\paperfont\scriptsize,line width=0.3pt,
    minimum width=42mm},
  arr/.style={->,line width=0.55pt},
  darr/.style={->,line width=0.55pt,densely dashed},
]
\node[box,fill=blue!6,minimum width=14mm] (wav) at (0,0) {Waveform};
\node[box,fill=blue!12,minimum width=18mm,right=5mm of wav]
  (enc) {Audio\\[-1pt]encoder};
\node[inner,fill=red!5,anchor=west]
  at ([xshift=14mm,yshift=5.5mm]enc.east) (ca)
  {Question-guided cross-attention};
\node[inner,fill=yellow!8,below=2.5mm of ca] (sa)
  {Scale-aware self-attention};
\node[inner,fill=green!6,below=2.5mm of sa] (ffn)
  {Feed-forward\;+\;residual};
\node[font=\paperfont\tiny,text=black!50,anchor=east]
  at ([xshift=-1mm]sa.east) {$w{=}2,\!4,\!8,\!16$};
\node[draw,densely dashed,rounded corners=3pt,
  inner xsep=5pt,inner ysep=4.5pt,
  fit=(ca)(sa)(ffn),line width=0.45pt] (rlbbox) {};
\node[font=\paperfont\footnotesize,anchor=south,yshift=1pt]
  at (rlbbox.north) {Recurrent Latent Block\;\textit{(shared weights)}};
\node[font=\paperfont\tiny,text=black!55,anchor=south west]
  at ([xshift=1mm,yshift=0.5mm]ca.north west) {$+\,\mathbf{e}_k$};
\draw[arr,black!55] (ca)--(sa);
\draw[arr,black!55] (sa)--(ffn);
\draw[darr,thick,black!65]
  ([xshift=3pt]ffn.south east)
  ..controls +(7mm,-4mm) and +(7mm,4mm)..
  node[right,font=\paperfont\scriptsize,xshift=0.5mm]{$\times\;K$}
  ([xshift=3pt]ca.north east);
\node[box,fill=orange!10,minimum width=22mm,
  below=8mm of rlbbox] (halt) {Halting unit};
\draw[arr] ([xshift=6mm]ffn.south)--([xshift=6mm]halt.north);
\node[font=\paperfont\tiny,text=orange!60!black,anchor=west]
  at ([xshift=1mm]halt.east) {stop if\;$R_k\!\ge\!1$};
\node[box,fill=red!6,minimum width=20mm,
  left=10mm of halt] (q) {Question\;$\mathbf{q}$};
\draw[arr] (q.north)--([xshift=-6mm]ca.south);
\draw[arr] (q)--(halt);
\draw[arr] (wav)--(enc);
\draw[arr] (enc.east)--
  node[above,font=\paperfont\scriptsize]{$\mathbf{H}^{(0)}$}
  (ca.west |- enc.east);
\node[box,fill=green!8,minimum width=18mm,
  right=10mm of rlbbox.east |- enc] (lm) {Language\\[-1pt]model};
\node[box,fill=green!14,minimum width=14mm,right=5mm of lm] (ans) {Answer};
\draw[arr] (ffn.east |- enc)--
  node[above,font=\paperfont\scriptsize]{$\mathbf{H}^{(K)}$} (lm);
\draw[arr] (lm)--(ans);
\end{tikzpicture}
\vspace{-3pt}
\caption{Overview of AdaLoop. The Recurrent Latent Block (dashed) applies
question-guided cross-attention, scale-aware self-attention with four window
sizes ($w$), and a feed-forward update. Shared parameters iterate $K$ times;
a halting unit conditioned on both the pooled audio state and the question
decides when to stop.}
\label{fig:overview}
\vspace{-5pt}
\end{figure*}

\vspace{-5pt}
\section{Adaptive-Depth Latent Reasoning}
\label{sec:method}
\vspace{-5pt}

AdaLoop replaces the linear projection that connects the audio encoder to the
language model with a recurrent module whose depth adapts to each input
(Fig.~\ref{fig:overview}). Given audio encoder features
$\mathbf{H}^{(0)}\!\in\!\mathbb{R}^{L\times d}$ and the question embedding
$\mathbf{q}$, AdaLoop applies $K$ iterations of a Recurrent Latent Block
(RLB), where $K$ is determined by a learned halting mechanism, then passes the
refined features $\mathbf{H}^{(K)}$ to the language model backbone.

\vspace{-3pt}
\subsection{Recurrent Latent Block}
\vspace{-3pt}

Each iteration $k\in\{1,\ldots,K\}$ applies three operations whose parameters
are shared across iterations; a learned iteration embedding
$\mathbf{e}_k\!\in\!\mathbb{R}^{d}$ added at the start of each pass breaks
the symmetry so that early and late iterations can specialize.

\textbf{Question-guided cross-attention.}
The block first lets the audio features attend to the question:
\begin{equation}
\mathbf{A}^{(k)} =
  \operatorname{CrossAttn}\!\bigl(\mathbf{H}^{(k-1)}\!+\mathbf{e}_k,\;\mathbf{q}\bigr)
  + \mathbf{H}^{(k-1)},
\end{equation}
where the query projection operates on
$\mathbf{H}^{(k-1)}\!+\mathbf{e}_k$ and the key and value projections
on~$\mathbf{q}$.

\textbf{Scale-aware self-attention.}
Fine-grained acoustic cues operate at different temporal scales: plosive onsets
span a few milliseconds, while pitch contours and rhythms develop over hundreds.
We partition the $M$ attention heads into $G{=}4$ groups and restrict each
group~$g$ to a local window of $w_g{=}2^{g+1}$ positions, following the
local attention principle of Longformer~\cite{longformer}, yielding receptive
fields that double from group to group. This lets one pass capture both fast
spectral transients and slow prosodic structure:
\begin{equation}
\mathbf{S}^{(k)} =
  \operatorname{ScaleAttn}\!\bigl(\mathbf{A}^{(k)}\bigr) + \mathbf{A}^{(k)}.
\end{equation}

\textbf{Feed-forward update.}
A standard feed-forward network~\cite{transformer} with a residual
connection~\cite{resnet} completes the
iteration:
\begin{equation}
\mathbf{H}^{(k)} =
  \mathbf{H}^{(k-1)}
  + \operatorname{FFN}\!\bigl(\operatorname{LN}(\mathbf{S}^{(k)})\bigr).
\end{equation}
The pre-norm residual~\cite{layernorm} connects directly to $\mathbf{H}^{(k-1)}$ rather than to
$\mathbf{S}^{(k)}$, ensuring that the identity is always an easy solution and
stabilizing training when many iterations are chained.

\vspace{-3pt}
\subsection{Adaptive Halting}
\vspace{-3pt}

A halting unit decides when the representation has been refined enough.
After each iteration it computes a scalar halting probability conditioned on
both the current state and the question:
\begin{equation}
p_k = \sigma\!\bigl(\mathbf{w}^{\!\top}[\,\overline{\mathbf{H}}^{(k)};\,
  \overline{\mathbf{q}}\,] + b\bigr),
\label{eq:halt}
\end{equation}
where $\overline{\mathbf{H}}^{(k)}$ and $\overline{\mathbf{q}}$ denote mean
pooling over the respective sequence lengths, $[\,\cdot\,;\cdot\,]$ is
concatenation, and $\sigma$ is the sigmoid function. Including the question
lets the halting unit learn that a language-identification query needs fewer
passes than a pitch-comparison query.

The cumulative halting signal is $R_k = \sum_{j=1}^{k}p_j$. Iteration stops at
$K = \min\{k : R_k \ge 1\}$, capped at $K_{\max}{=}8$. Following adaptive
computation time~\cite{graves2016adaptive}, the output is the weighted mixture
\begin{equation}
\mathbf{H}_{\mathrm{out}} =
  \sum_{k=1}^{K-1}p_k\,\mathbf{H}^{(k)}
  + r\,\mathbf{H}^{(K)},
  \quad r = 1 - R_{K-1},
\label{eq:act}
\end{equation}
which is differentiable with respect to the halting parameters, so the full
module can be trained end to end.

\vspace{-3pt}
\subsection{Training}
\vspace{-3pt}

The loss combines the standard next-token prediction objective $\mathcal{L}_{\text{task}}$
with a ponder cost that encourages the model to use fewer iterations when
possible:
\begin{equation}
\mathcal{L} =
  \mathcal{L}_{\text{task}}
  + \lambda\,\frac{1}{N}\sum_{i=1}^{N}K_i,
  \qquad \lambda = 0.01,
\label{eq:loss}
\end{equation}
where $K_i$ is the number of iterations for the $i$-th sample in a batch of
$N$. Training proceeds in two stages: first the RLB and halting unit are trained
for $1\,000$ steps with the audio encoder and language model frozen, then all
trainable components (RLB, halting unit, language model) are fine-tuned jointly
for $3\,000$ steps. The audio encoder stays frozen throughout.
Stage~1 uses a learning rate of $5{\times}10^{-5}$ and stage~2 drops to
$1{\times}10^{-5}$; both use the Adam optimizer~\cite{adam}, linear warmup over 100 steps, and cosine decay.
The training set consists of 50\,000 question--answer pairs drawn from the
verifiable audio QA families of EvoAudio~\cite{wang2026evoaudio}, covering
speech prosody (pitch, rate, and loudness comparisons), speakers and dialogue
(turn counting, speaker similarity), speech content (intent, emotion), sound
events (counting, ordering, identification), music (tempo, dynamics, instrument
recognition), and multi-clip scenes. Each answer follows from how the audio was
constructed or from labels the source recording already carries.
Source audio is drawn from LibriSpeech~\cite{librispeech},
FSD50K~\cite{fsd50k}, AudioSet~\cite{audioset}, AudioCaps~\cite{audiocaps},
and MELD~\cite{meld}, with synthesis by Qwen3-TTS~\cite{tts},
LeVo~\cite{levo}, and procedural mixing via Scaper~\cite{scaper}. No family
exceeds 20\% of the set, ensuring that the model encounters every question type
in each epoch. The RLB uses pre-norm attention with rotary position embeddings~\cite{rope}
truncated to each group's window. All experiments cap the audio input at 3\,000
tokens (approximately 30 seconds), truncating with a centered window when
necessary.

\begin{table*}[t]
  \centering
  \caption{Accuracy (\%) on three audio-understanding benchmarks.
  Avg.\ is the mean of the three All scores.
  SFT fine-tunes only the connector on the same data used by AdaLoop.
  Fixed-4 uses the same RLB architecture but with a fixed depth of four
  iterations and no halting mechanism.
  For MMAU-Pro, the three category columns do not cover every scored item,
  so All is not their average. \textbf{Bold} and \second{underline} mark
  the best and second best per backbone.}
  \label{tab:main}
  \centering
  \fontsize{9}{9.4}\selectfont
  \setlength{\tabcolsep}{1.5pt}
  \renewcommand{\arraystretch}{0.9}
  \arrayrulecolor{tblRule}
  \begin{tabular*}{\textwidth}{@{\extracolsep{\fill}}l*{12}{N}A@{}}
  \toprule
  & \multicolumn{3}{c}{\hd{MMSU $\uparrow$}} & \multicolumn{4}{c}{\hd{MMAU-Pro $\uparrow$}}
  & \multicolumn{5}{c}{\hd{MMAR $\uparrow$}} & \\
  \cmidrule(lr){2-4}\cmidrule(lr){5-8}\cmidrule(lr){9-13}
  \hd{Method} & \hd{Percep.} & \hd{Reason.} & \hd{All} & \hd{Speech} & \hd{Sound} & \hd{Music} & \hd{All} & \hd{Signal} & \hd{Percep.} & \hd{Seman.} & \hd{Culture} & \hd{All} & \hd{Avg. $\uparrow$} \\
  \midrule
  \rowcolor{tblBand}
  \multicolumn{14}{@{}l}{\grp{Qwen2.5-Omni-7B}} \\
  \addlinespace[1.3pt]
  Base & 44.5 & 79.4 & 61.4 & 57.8 & 45.3 & 64.0 & 56.4 & 55.8 & 54.0 & 67.2 & 58.9 & 60.2 & 59.3\phantom{\gain{0.0}} \\
  SFT & 47.2 & 80.1 & 63.2 & 58.9 & 46.8 & 65.1 & 57.4 & 57.0 & 55.2 & 67.5 & 59.4 & 60.8 & 60.5\gain{1.2} \\
  Fixed-4 & \second{49.8} & \second{80.6} & \second{64.8} & \second{59.5} & \second{47.5} & \second{65.8} & \second{58.0} & \second{57.9} & \second{55.8} & \second{67.7} & \second{60.0} & \second{61.3} & \second{61.4}\gain{2.1} \\
  \rowcolor{tblOurB}
  \ours{AdaLoop} & \best{53.1} & \best{81.7} & \best{67.0} & \best{60.8} & \best{49.1} & \best{66.5} & \best{59.4} & \best{59.5} & \best{56.9} & \best{68.1} & \best{61.2} & \best{62.8} & \best{63.1}\gain{3.8} \\
  \midrule
  \rowcolor{tblBand}
  \multicolumn{14}{@{}l}{\grp{Kimi-Audio-7B-Instruct}} \\
  \addlinespace[1.3pt]
  Base & 39.0 & 74.3 & 56.0 & 58.9 & 37.1 & 56.1 & 49.6 & 46.5 & 47.3 & 63.1 & 51.1 & 54.3 & 53.3\phantom{\gain{0.0}} \\
  SFT & 41.3 & 75.5 & 57.9 & 60.0 & 39.2 & 57.3 & 50.9 & 48.1 & 48.5 & 63.8 & 52.3 & 55.2 & 54.7\gain{1.4} \\
  Fixed-4 & \second{42.5} & \second{76.0} & \second{58.8} & \second{60.4} & \second{40.0} & \second{57.8} & \second{51.4} & \second{49.0} & \second{49.0} & \second{64.2} & \second{52.7} & \second{55.8} & \second{55.3}\gain{2.0} \\
  \rowcolor{tblOurB}
  \ours{AdaLoop} & \best{44.8} & \best{76.9} & \best{60.4} & \best{61.6} & \best{42.1} & \best{58.9} & \best{52.9} & \best{51.2} & \best{50.1} & \best{64.8} & \best{53.8} & \best{56.9} & \best{56.7}\gain{3.4} \\
  \midrule
  \rowcolor{tblBand}
  \multicolumn{14}{@{}l}{\grp{MiMo-Audio-7B-Instruct}} \\
  \addlinespace[1.3pt]
  Base & 46.9 & 73.1 & 59.6 & 62.6 & 38.3 & 64.3 & 54.9 & 44.2 & 55.9 & 69.2 & 62.4 & 61.8 & 58.8\phantom{\gain{0.0}} \\
  \rowcolor{tblOurB}
  \ours{AdaLoop} & \best{51.8} & \best{76.0} & \best{63.5} & \best{64.5} & \best{42.5} & \best{66.3} & \best{57.8} & \best{49.5} & \best{58.6} & \best{70.5} & \best{64.0} & \best{63.7} & \best{61.7}\gain{2.9} \\
  \bottomrule
  \end{tabular*}
  \arrayrulecolor{black}
  \vspace{-5pt}
\end{table*}

\vspace{-5pt}
\section{Experiments}
\label{sec:experiments}
\vspace{-5pt}

\vspace{-3pt}
\subsection{Setup}
\vspace{-3pt}

We evaluate AdaLoop on three LALMs with different audio encoders and language
backbones: Qwen2.5-Omni-7B~\cite{qwen25}, Kimi-Audio-7B-Instruct~\cite{kimi},
and MiMo-Audio-7B-Instruct~\cite{mimo}. These models span encoder
families from Whisper-based~\cite{whisper} to self-supervised~\cite{hubert}
and conformer~\cite{conformer}, paired with language backbones derived from
Qwen2~\cite{qwen2}, LLaMA~\cite{llama2}, and
DeepSeek~\cite{deepseekv3}. Recent models including Audio
Flamingo~\cite{af2,af3} and MiniCPM-o~\cite{minicpm} could also benefit.
For each model, the RLB matches the
hidden dimension of the original connector. The RLB has 8 attention heads split
into 4 scale groups with windows $\{2,4,8,16\}$, one feed-forward layer of
expansion ratio~4, and $K_{\max}{=}8$ iteration embeddings. This adds 2.4\% to
2.9\% of the base model's parameter count. Training runs on 4 NVIDIA A100-80G
GPUs with per-GPU batch size~2 and gradient accumulation over 4 steps
(effective batch 32).

We compare four methods: (1)~\textbf{Base}---the released checkpoint;
(2)~\textbf{SFT}---the same 50K training set is used to fine-tune only the
original connector via supervised learning; (3)~\textbf{Fixed-4}---the RLB
architecture with a constant depth $K{=}4$ and no halting mechanism;
(4)~\textbf{AdaLoop}---our full adaptive-depth module. For MiMo-Audio we
report Base and AdaLoop only, as its connector architecture does not admit a
drop-in replacement for the intermediate methods.

We evaluate on MMSU~\cite{mmsu}, MMAU-Pro~\cite{mmaupro}, and
MMAR~\cite{mmar}. Together these benchmarks cover spoken perception and
reasoning (MMSU), speech, sound, and music understanding (MMAU-Pro), and
signal-level, perceptual, semantic, and cultural reasoning about audio (MMAR).
For MMAU-Pro we use the same closed nonspatial subset as prior
work~\cite{wang2026evoaudio,omnir1}; invalid predictions count as incorrect.

\subsection{Main Results}

Table~\ref{tab:main} shows a consistent ordering across all three backbones.
AdaLoop achieves the highest average and the highest score on every individual
benchmark for every model. Relative to Base, it raises the average by 3.8
points on Qwen2.5-Omni, 3.4 on Kimi-Audio, and 2.9 on MiMo-Audio. The
gains are broad: on Qwen2.5-Omni, every one of the twelve subcategories
improves, with the largest single jump on MMSU perception (+8.6 points).

The progression from SFT through Fixed-4 to AdaLoop isolates the sources of
improvement. SFT uses the same training data but only replaces the connector
weights, gaining 1.2 points on Qwen2.5-Omni. Moving to Fixed-4 adds iterative
refinement and another 0.9 points. AdaLoop then introduces adaptive depth for a
further 1.7 points. The largest step comes from the adaptive mechanism, not
from the training data or the recurrent architecture alone. The pattern mirrors
findings in language modelling, where reinforcement
learning~\cite{ppo,dpo,rlhf,grpo,dapo} and chain-of-thought
reasoning~\cite{cot} both improve by giving the model more targeted
computation.

Perception-heavy categories benefit the most. On Qwen2.5-Omni, MMSU perception
rises by 8.6 points while MMAR semantic rises by only 0.9. This pattern holds
across backbones: Kimi-Audio's MMAU-Pro sound score (the weakest subcategory)
improves by 5.0 points, and MiMo-Audio's MMAR signal score rises by 5.3.
Semantic subcategories, where a coarse representation often suffices, see
smaller but consistently positive gains.

\subsection{Learned Reasoning Depth}

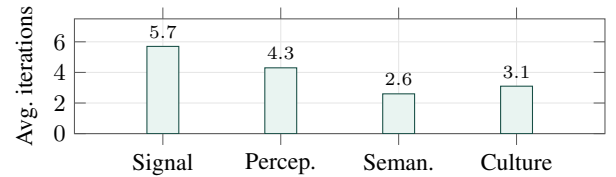
\begin{figure}[t]
\centering
\begin{tikzpicture}
\begin{axis}[
  ybar, bar width=12pt,
  width=\columnwidth, height=3.1cm,
  ylabel style={font=\paperfont\small},
  ylabel={Avg.\ iterations},
  symbolic x coords={Signal,Percep.,Seman.,Culture},
  xtick=data,
  tick label style={font=\paperfont\small},
  ymin=0, ymax=7.5,
  ytick={0,2,4,6},
  nodes near coords,
  every node near coord/.append style={font=\paperfont\scriptsize},
  axis line style={black!65}, tick style={black!65},
  grid=major, grid style={black!10},
  enlarge x limits=0.25,
]
\addplot[fill=tblOurA,draw=tblOurInk] coordinates
  {(Signal,5.7) (Percep.,4.3) (Seman.,2.6) (Culture,3.1)};
\end{axis}
\end{tikzpicture}
\vspace{-6pt}
\caption{Average AdaLoop iterations by MMAR category on Qwen2.5-Omni-7B.
Signal-level tasks trigger the deepest reasoning.}
\label{fig:depth}
\vspace{-8pt}
\end{figure}

Fig.~\ref{fig:depth} breaks down the average iteration count by MMAR
category on Qwen2.5-Omni. Signal-level questions---pitch comparisons,
loudness ordering, rate estimation---receive 5.7 iterations on average, nearly
triple the 2.6 used for semantic questions about content or intent. Perceptual
tasks (speaker identity, emotion) sit at 4.3, and cultural reasoning at 3.1.
Across all benchmarks the average is 3.8 iterations, ranging from a single pass
(9.3\% of items) to the maximum eight (4.2\%). These distributions are
consistent across backbones: Kimi-Audio averages 4.2 iterations (its weaker
perception starts from a harder position), and MiMo-Audio averages 3.9.

The depth ordering matches the nature of each task. Signal-level judgments
require precise analysis of temporal and spectral patterns that benefit from
repeated attention at different scales. Semantic questions, by contrast, can
often be resolved from a single high-level embedding that captures the gist
of the utterance. The fact that the model discovers this allocation from the
ponder cost alone, without any explicit depth labels, validates the adaptive
design.

Depth distributions also vary across backbones: Kimi-Audio averages 4.2
iterations (its weaker perception needs more refinement) and MiMo-Audio
averages 3.9, reaching 5.4 on signal-level tasks where its base accuracy is
the lowest of the three.

\vspace{-3pt}
\subsection{Ablations}
\vspace{-3pt}

\begin{table}[t]
\caption{Ablations on Qwen2.5-Omni-7B. Each row changes one component of
AdaLoop while keeping the training budget fixed. Base model Avg.\ is 59.3\%.}
\label{tab:ablation}
\centering
\arrayrulecolor{tblRule}
\paperfont
\setlength{\tabcolsep}{3.5pt}
\renewcommand{\arraystretch}{1.05}
\begin{tabularx}{\columnwidth}{X>{\centering\arraybackslash}p{3.8em}
  >{\centering\arraybackslash}p{4.6em}>{\centering\arraybackslash}p{3.6em}A}
\toprule
\hd{Variant} & \hd{MMSU} & \hd{MMAU-P} & \hd{MMAR} & \hd{Avg.\ (\%)} \\
\midrule
\rowcolor{tblOurB}
\ours{AdaLoop} & 67.0 & 59.4 & 62.8 & 63.1 \\
Fixed depth ($K{=}4$) & 64.8 & 58.0 & 61.3 & 61.4 \\
No scale-aware attn & 65.4 & 58.5 & 61.8 & 61.9 \\
No question in halt & 65.0 & 58.1 & 61.5 & 61.5 \\
No ponder cost & 65.8 & 58.7 & 62.3 & 62.3 \\
$K_{\max} = 4$ & 65.9 & 58.8 & 62.2 & 62.3 \\
\bottomrule
\end{tabularx}
\arrayrulecolor{black}
\vspace{-8pt}
\end{table}

Table~\ref{tab:ablation} isolates the contribution of each component.
Replacing adaptive halting with a fixed depth of four iterations causes the
largest single drop ($-1.7$ points), confirming that the depth adaptation is the
most valuable ingredient. Removing the question from the halting
input~(\ref{eq:halt}) costs 1.6 points: without question conditioning, the
model must decide depth from the audio alone and cannot distinguish an easy
question about a hard audio clip from a hard question about the same clip.
Scale-aware attention contributes 1.2 points; the uniform-window alternative
still benefits from recurrence but cannot simultaneously capture fast and slow
acoustic patterns. Removing the ponder cost ($\lambda{=}0$) and capping
$K_{\max}$ at~4 each cost 0.8~points. Without the ponder cost the model
defaults to deeper processing even for trivial queries, wasting capacity and
introducing noise from unnecessary iterations. The $K_{\max}$ ablation shows
that eight iterations provide headroom the model actually uses: the few
questions that reach eight passes include the hardest signal-level comparisons.

\vspace{-3pt}
\section{Conclusion}
\label{sec:conclusion}
\vspace{-3pt}

We introduced AdaLoop, a lightweight recurrent module that learns how many
latent refinement steps each audio question requires. It improves three
architecturally distinct LALMs on three benchmarks without modifying their
encoders or language backbones, and the gains concentrate on perception-heavy
tasks where deeper iterative reasoning is most needed. What limits AdaLoop
is the ceiling on tasks where a single coarse pass already suffices;
combining adaptive depth with richer training curricula~\cite{wang2026evoaudio,azr}
and parameter-efficient adaptation~\cite{lora} is a natural next step.

\balance
\paperfont\sloppy

\bibliographystyle{IEEEbib_etal}
\bibliography{refs}
\end{document}